\documentclass[prl,twocolumn]{revtex4}

\usepackage{graphicx}
\usepackage{dcolumn}
\usepackage{bm}
\usepackage{amsfonts,amsmath,amssymb}

\newcommand{\cN}{{\cal N}}

\newcommand{\0}{{(0)}}
\newcommand{\1}{{(1)}}
\newcommand{\2}{{(2)}}
\newcommand{\3}{{(3)}}
\newcommand{\4}{{(4)}}

\newcommand{\6}{{(6)}}

\allowdisplaybreaks

\begin{document}

\begin{flushright}
NIKHEF-2015-011 \quad CPHT-RR015.0415
\end{flushright}

\vspace{-2cm}

\title{The string origin of dyonic $\cN=8$ supergravity and its Chern-Simons duals}

\author{Adolfo Guarino$^{1}$, Daniel L. Jafferis$^2$ and Oscar
Varela$^{2,3}$}

%\email{aguarino@nikhef.nl, jafferis@physics.harvard.edu, ovarela@physics.harvard.edu}

\affiliation{$^1$Nikhef, Science Park 105, 1098 XG Amsterdam, The Netherlands
\\
$^2$Center for the Fundamental Laws of Nature, Harvard University, Cambridge, MA 02138, USA 
\\
$^3$\mbox{Centre de Physique Th\'eorique, Ecole Polytechnique, CNRS UMR 7644, 91128 Palaiseau Cedex, France} }

\begin{abstract}
We clarify the higher-dimensional origin of a class of dyonic gaugings of $D=4$ $\cN=8$ supergravity recently discovered, when the gauge group is chosen to be ISO(7). This dyonically-gauged maximal supergravity arises from consistent truncation of massive IIA supergravity on $S^6$, and its magnetic coupling constant descends directly from the Romans mass. The critical points of the supergravity uplift to new AdS$_4$ massive type IIA vacua. We identify the corresponding CFT$_3$ duals as super-Chern-Simons-matter theories with simple gauge group SU$(N)$ and level $k$ given by the Romans mass. In particular, we find a critical point that uplifts to the first explicit $\cN=2$ AdS$_4$ massive IIA background. We compute its free energy and that of the candidate dual Chern-Simons theory by localisation to a solvable matrix model, and find perfect agreement. This provides the first AdS$_4$/CFT$_3$ precision match in massive type IIA string theory. 
\end{abstract}

\pacs{}

\maketitle

\vspace{-20cm}

\noindent \textbf{Introduction}. Supergravity theories, the supersymmetric extensions of General Relativity, come in two varieties: gauged and ungauged. The former, unlike the latter, typically include non-Abelian gauge groups and a scalar potential. Gauged supergravities whose scalar potentials have supersymmetric anti-de-Sitter (AdS) critical points are particularly interesting since they can provide insights into the strong coupling behaviour of superconformal field theories (CFTs) through the AdS/CFT correspondence \cite{Maldacena:1997re}. The relevant gauged supergravities should arise as consistently truncated compactifications of $D=10$ or $D=11$ supergravity. Only if consistency holds will their AdS vacua (and any other solution) uplift to string or M-theory backgrounds on which AdS/CFT can be formulated precisely.  

Consider, for instance, the SO(8) gauging of the maximally supersymmetric, $\cN=8$, supergravity in $D=4$ \cite{de Wit:1982ig}. It arises from consistent truncation of M-theory on $S^7$ \cite{de Wit:1986iy,deWit:2013ija}. All its supersymmetric vacua uplift to AdS$_4 \times S^7$ M-theory backgrounds, some of which have known field theory duals. For example, the central critical point uplifts to the Freund-Rubin solution of $D=11$ supergravity, which is dual to the ABJM superconformal field theory \cite{Aharony:2008ug}:  a super-Chern-Simons theory with non-simple  gauge group U$(N) \times $U$(N)$ at (low) levels $k$ and $-k$. The SO(8) gauging of \cite{de Wit:1982ig} is purely electric, in the sense that it only involves the vectors that appear in the lagrangian, and not their magnetic duals. It has been recently pointed out that, more generally, $D=4$ $\cN=8$  gauged supergravities often admit dyonic gaugings \cite{Dall'Agata:2012bb,Dall'Agata:2014ita}. These are characterised by a dimensionless parameter, either continuous or discrete, that determines the linear combination of electric and magnetic vectors, in the adjoint of the gauge group, that participate in the gauging. This parameter shows up in the couplings of the gauged supergravity, particularly the scalar potential.

The questions arise: do these $\cN=8$ dyonic gaugings enjoy a string or M-theory origin, or are they just a four-dimensional artifact? And, closely related for supergravities with supersymmetric AdS vacua, are these dual to any three-dimensional CFTs? In this note we show that these questions have precise answers for the dyonic gauging of a group closely related to SO(8): its contraction $\textrm{ISO}(7) = \textrm{SO}(7) \ltimes \mathbb{R}^7$. We find that ISO(7)-dyonically-gauged $\cN=8$ supergravity arises as a consistent truncation of massive type IIA supergravity \cite{Romans:1985tz} on the six-sphere, with the magnetic coupling constant identified upon reduction with the Romans mass. This gauged supergravity has AdS critical points that uplift to new AdS$_4 \times S^6$ backgrounds, with deformed metrics on the $S^6$ and various amounts of supersymmetry. We also give quantitative evidence that massive IIA string theory on these backgrounds is dual to the simplest possible type of superconformal Chern-Simons theories: those, first considered by Schwarz \cite{Schwarz:2004yj} as potentially relevant for holography, with a simple gauge group SU$(N)$, adjoint matter and level $k$. As anticipated in \cite{Schwarz:2004yj} (see also \cite{Gaiotto:2009mv}), the level coincides with the quantised Romans mass. The $D=4$ magnetic coupling $m$, the $D=10$ Romans mass $\hat F_\0$ and the level $k$ of the CFT$_3$ duals are thus related by
\\[-13pt]
\begin{eqnarray} \label{QuantisationRomansMass}
 m = \hat F_\0 = k /(2\pi\ell_s) \; ,
\end{eqnarray}
\\[-13pt]
where $\ell_s = \sqrt{\alpha^\prime}$ is the string length.

\vspace{2pt}

\noindent \textbf{Dyonic ISO(7)-gauged supergravity}. The Romans mass is known to induce magnetic gaugings and mass terms for the NS two-form in $\cN=2$ compactifications of massive IIA on Calabi-Yau with fluxes \cite{Polchinski:1995sm,Louis:2002ny}. Non-semisimple gaugings also occur frequently in this context. Our construction can thus be regarded as an $\cN=8$ extension of those $\cN=2$ models. Magnetic couplings and non-trivial tensors in four dimensions come hand-in-hand, and the embedding tensor formalism \cite{de Wit:2007mt}, that we use, naturally incorporates both systematically.

The  $\cN=8$ family of ISO(7) gaugings is characterised completely by an embedding tensor $\Theta_{\mathbb{M}}{}^\alpha$ of the form \cite{DallAgata:2011aa}
\begin{equation}
\label{Theta_Def}
\Theta_{[AB]}{}^C{}_D = 2 \, \delta_{[A}^{C} \theta_{B]D}
\hspace{2mm}  \textrm{ , } \hspace{2mm}
{\Theta^{[AB]C}}_{D}  = 2 \, \delta^{[A}_{D} \xi^{B]C} \  .
\end{equation}
We have split the adjoint index $\alpha$ of SL(8) and fundamental index $\mathbb{M}$ of E$_{7(7)}$ into fundamental SL(8) indices $A= 1 , \ldots, 8$, and have defined
\begin{equation}
\label{theta_xi_matrices}
\theta= g \, \textrm{diag}(\mathbb{I}_{7},0)
\; , \quad
\xi= m \, \textrm{diag}(0_{7},1) \ ,
\end{equation}
with $g$ and $m$ the electric and magnetic coupling constants. The dyonically-gauging parameter mentioned in the introduction is simply the ratio $c =m/g$. For $g \neq 0$, this family of ISO(7) gaugings is discrete. It contains, in fact, only two members \cite{Dall'Agata:2014ita}: the purely electric case $m =0$ constructed long ago \cite{Hull:1984yy}, and the $m \neq 0$ case (all $m \neq 0$ supergravities happen to be equivalent \cite{Dall'Agata:2014ita}). This form of the embedding tensor implies that the SO(7) subgroup of ISO(7) is gauged electrically only, while the seven translations are gauged dyonically.

Using (\ref{Theta_Def}), (\ref{theta_xi_matrices}) in the general formalism of \cite{de Wit:2007mt}, we have constructed the bosonic sector of the $\cN=8$ theory \cite{Guarino:2015qaa}. This contains the 70 scalars of E$_{7(7)}/$SU(8), which can be packed in the symmetric matrix ${\cal M}_{\mathbb{M} \mathbb{N} }$; the ISO(7) electric (and magnetic) vectors $A^{IJ}$, $A^I$ (and $\tilde{A}_{IJ}$, $\tilde{A}_I$), $I=1, \ldots, 7$, with field strengths \mbox{${\cal H}_\2^{IJ}$, ${\cal H}_\2^{I}$}; and other higher-rank tensors, including two-forms $B^{I}$, required by the vector-tensor hierarchy. The electrically-gauged SO(7) rotations lead to a conventional  ${\cal H}^{IJ}_\2$,
whereas the  field strengths of the dyonically-gauged $\mathbb{R}^7$ translations,%
\begin{equation}
\label{HR7D=4}
 {\cal H}^I_\2 = d A^I -g \delta_{JK} A^{IJ} \wedge A^K +\tfrac12 m A^{IJ} \wedge \tilde A_J +m \,  B^I  ,
\end{equation}
 include couplings to the magnetic vectors and to $B^I$. The two-forms acquire a topological mass
$g m \, \delta_{IJ} B^I \wedge B^J$,
similar to that in  \cite{Louis:2002ny}. Finally, the scalar covariant derivatives develop dyonic couplings, as expected, and the scalar potential features the terms in $g^2$ of the purely electric gauging \cite{Hull:1984yy} plus new $gm$ and $m^2$ terms.

It is often insightful to consider smaller sectors of the $\cN=8$ theory. A useful one is obtained by truncating bosons and fermions to the singlets under the SU$(3)$ subgroups of the gauge group ISO(7) and R-symmetry group SU$(8)$, respectively. This truncation  results in an $\cN=2$ subsector, including one vector multiplet and one hypermultiplet, with a $\textrm{U}(1) \times \textrm{SO}(1,1)$ gauging in the hyper sector. A further consistent truncation of this sector retains only the metric and the three scalars neutral under the gauge group, and is described by the Lagrangian
\begin{equation} \label{SU3U1sectorLag}
e^{-1} {\cal L} = R -2 (\partial \phi)^2  -\tfrac{3}{2} ( \partial \varphi )^2  - \tfrac{3}{2} \,e^{2\varphi} (\partial \chi )^2 -V \ ,
\end{equation}
where the scalar potential reads
\begin{eqnarray} \label{PotSU3U1Geom}
&&  V = \tfrac12  g^2  \Big(
e^{4\phi-3\varphi} \big( 1+ e^{2\varphi} \chi^2 \big)^3
-12 e^{2\phi-\varphi} \big( 1+ e^{2\varphi} \chi^2 \big)  \nonumber  \\
&& \qquad\qquad \;  -24 e^{\varphi} \Big)
 -g m \,e^{4\phi +3\varphi}  \chi^3
 +\tfrac12 m^2 e^{4\phi +3\varphi}  \, .
\end{eqnarray}
For $ g m\neq 0$, this potential has three AdS critical points. Two of them have already been predicted \cite{DallAgata:2011aa} by a different method \cite{Dibitetto:2011gm,DallAgata:2011aa}: they are non-supersymmetric, unstable, and respectively preserve SO(7) and SO(6) symmetry when embedded in the full $\cN=8$ ISO(7) theory. Curiously \cite{DallAgata:2011aa}, their mass spectra coincide with those of the SO(7)$_\pm$ and SU$(4)_-$ points of the SO(8) gauging. In addition, we find a new critical point located  at
\begin{eqnarray} \label{SU3U1pointSU3U1sector}
e^{6 \varphi}  = \tfrac{64}{27} \, g^2 m^{-2} \; , \; e^{6 \phi} = 8 \, g^2 m^{-2} \; , \; \chi^3 =  -\tfrac18 \, g^{-1} m  \ .
\end{eqnarray}
When embedded in the full $\cN=8$ ISO(7) theory, this point preserves $\cN=2$ supersymmetry and SU(3)$\times $U(1) bosonic symmetry. We have calculated its mass spectrum: again, it coincides with the spectrum  \cite{Klebanov:2008vq} of the $\cN=2$ $\textrm{SU}(3)\times \textrm{U}(1)$ point of the SO(8) gauging.

\vspace{2pt}

\noindent \textbf{Consistent truncation from massive IIA}. We have built the $D=10$ embedding of the full ISO(7) theory using a similar strategy employed to embed the electric $D=4$ SO(8) gauging into $D=11$ \cite{de Wit:1986iy,deWit:2013ija} or the $D=5$ SO(6) gauging into type IIB \cite{Ciceri:2014wya}. Firstly, redefinitions of the IIA fields are performed that leave only a subgroup SO$(1,3)$ of the full SO$(1,9)$ local Lorentz symmetry manifest. Secondly, the supersymmetry variations of these redefined fields are manipulated so that they conform to the E$_{7(7)}$-covariant vector-tensor hierarchy, and `generalised vielbeine' can be read off. Finally, an ansatz is proposed that relates the generalised vielbeine and the hierarchy-compatible vectors and tensors with the $D=4$ coset representative and vectors and tensors of the ISO(7) theory, together with geometrical data from  $S^6$. We have verified the consistency of this ansatz at the level of the supersymmetry variations: all $S^6$ data drop out, yielding the variations of the $D=4$ ISO(7) theory. 

Here we will only give the final result. Further details of this long analysis will be presented separately  \cite{Guarino:2015vca}. Leaving also for \cite{Guarino:2015vca} the rather long expression for $\hat A_\3$, the exact, non-linear consistent embedding reads, in the type IIA Einstein frame conventions of \cite{Cvetic:1999un},
\begin{equation}
\label{KKEmbedding}
{\setlength\arraycolsep{0.90pt}
\begin{array}{llll}
d\hat{s}_{10}^2 &=&  \Delta^{-1} \, ds_4^2  \, + g_{mn}  \, Dy^m \, Dy^n  & , \\[3pt]
e^{-\frac32 \hat  \phi} &=& -g^{mn} A_m A_n +  \Delta \,\mu_I \, \mu_J \,  {\cal M}^{I8 \, J8} & , \\[3pt]
\hat B_\2 &=& -\mu_I \big( B^I  +\tfrac12 A^{IJ} \wedge \tilde{A}_J \big)   - g^{-1}  \, \tilde A_{I}   \wedge D \mu^I   \\[3pt]
&& + \tfrac12  B_{mn} \,  Dy^m \wedge Dy^n & , \\[3pt]
\hat A_\1 &=& - \mu_I  \, A^{I}  + A_{m }  \, Dy^m & ,
\end{array}
}
\end{equation}
with $\Delta^2 = (\det \, g_{mn})/( \det \, \mathring g_{mn} )$, where $ \mathring g_{mn}$ is the round, SO(7)-symmetric metric on $S^6$. The $\mu^I$ parameterise $S^6$ as the locus $\mu_I \mu^I =1$ in $\mathbb{R}^7$, and $y^m$, $m=1, \ldots , 6$, are the $S^6$ angles. These have covariant derivatives
\begin{eqnarray}
\label{DyDmu}
Dy^m \equiv  dy^m + \tfrac12 g K_{IJ}^m A^{IJ}
\,\,\,\,,\,\,\,\,
D\mu^I  \equiv d \mu^I -g  A^{IJ} \mu_J ,
\end{eqnarray}
with $K^m_{IJ} = 2 g^{-2} \mathring{g}^{mn} \mu_{[I} \partial_n \mu_{J]}$  the  Killing vectors of $ \mathring{g}_{mn}$. Finally, the internal (inverse) metric and forms in (\ref{KKEmbedding}) are given in terms of SL(7)-covariant blocks of the $D=4$ scalar matrix  ${\cal M}_{\mathbb{M} \mathbb{N} }$ and $S^6$ quantities as
\begin{equation}
\label{KKmetric2}
\begin{array}{lll}
g^{mn} &=&  \tfrac14  g^2  \, \Delta \,    K^m_{IJ} \, K^n_{KL} \,  {\cal M}^{IJ \, KL} \ ,\\[5pt]
A_m &=&   \tfrac12 \, g \,  \Delta \, g_{mn} \, K^n_{IJ} \,  \mu_K \, {\cal M}^{IJ \, K8} \  ,\\[5pt]
B_{mn} &=& -  \tfrac12   \,   \Delta \, g_{mp} \, K^p_{IJ} \, \partial_{n} \mu^K \, {\cal M}^{IJ}{}_{K8} \ ,\\[5pt]
A_{mnp} &=&  A_m B_{np}  +   \tfrac18  \, g \,      \Delta  \, g_{mq} \, K^q_{IJ}  \, K_{np}^{KL}  {\cal M}^{IJ}{}_{KL}   .
\end{array}
\end{equation}
We have included the expression for the internal components of $\hat A_\3$, and have defined $K_{mn}^{IJ} = 4 g^{-2} \partial_{[m} \mu^I \partial_{n]} \mu^J$.

The electric coupling $g$ appears explicitly in these formulae, whereas the magnetic coupling $m$ does not. As usual in spherical reductions, $g$ becomes identified with the $S^6$ inverse  radius and must be non-vanishing for (\ref{KKEmbedding})--(\ref{KKmetric2}) to be well defined. In order to see that $m$ descends from the Romans mass $\hat F_\0$, we compute from (\ref{KKEmbedding}) the RR field strength \cite{Cvetic:1999un} $\hat F_\2 = d\hat A_\1 +  \hat F_\0 \, \hat B_\2$, and similarly for the other forms. We obtain
$\hat F_\2 = - \mu_I {\cal H}^I_\2 + \ldots $,
with
\\[-16pt]
\begin{equation} \label{HR7D=10}
 {\cal H}^I_\2 \equiv d A^I -g \delta_{JK} A^{IJ} \wedge A^K +\tfrac12 \hat F_\0 A^{IJ} \wedge \tilde A_J   + \hat F_\0 \,  B^I  
 \end{equation}
and the dots denoting scalar-dependent terms. This expression coincides with the $D=4$ field strength (\ref{HR7D=4}) provided the first identification in (\ref{QuantisationRomansMass}) holds. We have investigated further the correspondence between $m$ and $\hat F_0$ in the SU(3) and other invariant sectors, where explicit expressions for the  covariant derivatives and potential become available. Perfect matching between $D=4$ and $D=10$ is always found via (\ref{KKEmbedding})--(\ref{KKmetric2}) provided  (\ref{QuantisationRomansMass}) holds.

Being independent of $m$, the formulae (\ref{KKEmbedding})--(\ref{KKmetric2}) hold for $m=0$ as well, and thus also realise the embedding of the electric ISO(7) gauging \cite{Hull:1984yy} in massless IIA argued in \cite{Hull:1988jw}. Also, the discreteness \cite{Dall'Agata:2014ita} of the family of dyonic  ISO(7) gaugings can be understood directly in $D=10$: all non-vanishing values of $\hat F_\0$ are classically equivalent.

\vspace{2pt}

\noindent \textbf{A new $\cN=2$ AdS$_4$ massive IIA solution}. The consistent embedding (\ref{KKEmbedding})--(\ref{KKmetric2}) allows one to uplift any solution of ISO(7) supergravity to massive IIA, preserving supersymmetry in the process if present. We have employed these formulae to uplift the critical point (\ref{SU3U1pointSU3U1sector}) to obtain the first explicit, analytic $\cN=2$ AdS$_4$ solution of massive IIA supergravity we are aware of. In the IIA conventions of \cite{Cvetic:1999un}, the Einstein frame solution reads
\\[-15pt]
{\setlength\arraycolsep{0pt}
\begin{eqnarray} \label{SU3U1lIIASolutionRescaled}
&& d \hat{s}_{10}^2 =  L^2  \big( 3 + \cos 2\alpha \big)^{1/2} \big( 5 + \cos 2\alpha \big)^{1/8}   \Big[ \,  ds^2(\textrm{AdS}_4)   \nonumber \\
&& \qquad + \frac32 \,  d\alpha^2 +  \frac{ 6 \sin^2 \alpha}{ 3 + \cos 2\alpha } \,ds^2 ( \mathbb{CP}^2 ) % 
+   \frac{ 9 \sin^2 \alpha}{ 5 + \cos 2\alpha } \, \bm{\eta}^2  \Big] \; , \nonumber  \\
&& e^{\hat \phi} =  e^{\phi_0} \frac{  \big( 5 + \cos 2\alpha \big)^{3/4} }{ 3 + \cos 2\alpha }  \;  , \nonumber  \\
&&  L^{-2} e^{-\frac12 \phi_0}  \hat H_\3 =   24 \sqrt{2} \, \frac{  \sin^3 \alpha }{   \big( 3 + \cos 2\alpha \big)^2   }  \ \bm{J}  \wedge d\alpha  \; ,
\nonumber \\
&& L^{-3} e^{\frac14 \phi_0}   \hat F_\4 =  6 \, \textrm{vol} ( \textrm{AdS}_4 )  \nonumber \\
&& \qquad  \qquad  +  12\sqrt{3} \, \frac{  7 +  3\cos 2\alpha }{   \big( 3 + \cos 2\alpha \big)^2   } \, \sin^4 \alpha \ \textrm{vol} (  \mathbb{CP}^2)  \nonumber \\
&& \qquad  \qquad + 18\sqrt{3}  \, \frac{  (9 +  \cos 2\alpha) \sin^3 \alpha \cos \alpha }{   \big( 3 + \cos 2\alpha \big)  \big( 5 + \cos 2\alpha \big)   }  \ \bm{J}  \wedge d\alpha \wedge \bm{\eta} \; ,
\nonumber \\
&&  L^{-1} e^{\frac34 \phi_0}  \hat F_\2 =   -4  \sqrt{6} \,  \frac{  \sin^2 \alpha \cos \alpha  }{   \big( 3 + \cos 2\alpha \big)    \big( 5 + \cos 2\alpha \big)   }  \ \bm{J}    \nonumber \\
&& \qquad\qquad \qquad \quad -3  \sqrt{6} \,  \frac{ \big( 3 - \cos 2\alpha \big)  }{   \big( 5 + \cos 2\alpha \big)^2   }   \, \sin \alpha \ d\alpha \wedge \bm{\eta}   \; ,
\end{eqnarray}
 }with $L^2 \equiv 2^{-\frac{5}{8}} \, 3^{-1} \, g^{-\frac{25}{12}} \,  m^{\frac{1}{12}}$ and $e^{\phi_0} \equiv  2^{\frac{1}{4}} \, g^{\frac{5}{6}} \,  m^{-\frac{5}{6}}$, and the Romans mass given by the first equality in (\ref{QuantisationRomansMass}), namely, $\hat F_\0 = 3^{-\frac12} L^{-1} e^{-\frac54 \phi_0}$. As a check on our uplifting formulae, we have explicitly verified that (\ref{SU3U1lIIASolutionRescaled}) solves all the massive IIA field equations. 

The metrics on AdS$_4$ and Fubini-Study on $\mathbb{CP}^2$ are normalised so that the Ricci tensor equals $-3$ and $6$ times the metric, respectively. The angle $0 \leq \alpha \leq \pi$ locally foliates $S^6$ with $S^5$ leaves regarded as Hopf fibrations over $\mathbb{CP}^2$, with fibers squashed as a function of $\alpha$. Also, $\bm{J}$ is the K\"ahler form of $\mathbb{CP}^2$ and $\bm{\eta} = d\psi + \sigma$, with $0 \leq \psi \leq 2\pi$ a coordinate along the fiber and $d\sigma = 2\bm{J}$. The local internal metric can be alternatively regarded as one on an $S^2$ bundle over $\mathbb{CP}^2$, with $S^2$ fibers parameterised by $(\alpha , \psi)$ and $S^6$ topology for the total space. The local geometry extends globally over $S^6$ in a smooth manner. The vector $\partial_\psi$ is Killing, and also a symmetry of the supergravity forms, so that the full solution exhibits a cohomogeneity-one $\textrm{SU(3)} \times \textrm{U}(1)$ symmetry. The $\cN=2$ supersymmetry manifests itself in the form of a local SU(2)-structure, or global $\textrm{SU}(3) \times \textrm{SU}(3)$-structure, of the type discussed in \cite{Petrini:2009ur,Lust:2009mb}. Finally, a wider class of $\cN=2$ solutions with other topologies or possibly singular may be obtained from (\ref{SU3U1lIIASolutionRescaled}) by replacing $\mathbb{CP}^2$ with any positive-curvature K\"ahler-Einstein manifold or orbifold.

The supergravity solution  (\ref{SU3U1lIIASolutionRescaled}) also extends to a well-defined string background upon flux quantisation. On our topologically $S^6$ solution, flux quantisation conditions can only be imposed on $\hat F_\0$ and $\hat F_\6$. These are respectively given by the second relation in (\ref{QuantisationRomansMass}) and
{\setlength\arraycolsep{-5pt}
\begin{eqnarray} \label{quantF6}
&&  \frac{ -1  }{(2\pi \ell_s )^5 } \,  \int_{S^6}  e^{\frac12 \hat \phi} \   \hat{*} \hat F_4 + \hat B_2 \wedge d\hat A_3 + \frac16 \hat F_0 (\hat B_2)^3  = N \; ,
\end{eqnarray}
}with $N$ integer and $ (\hat B_2)^3 = \hat B_2 \wedge \hat B_2 \wedge \hat B_2$ .
From an explicit evaluation of this integral using (\ref{SU3U1lIIASolutionRescaled}) and from  (\ref{QuantisationRomansMass}), it is straightforward to solve for the classical parameters $L$, $e^{\phi_0}$ (or $g$, $m$) in terms of the quantum numbers $N$, $k$.

For later comparison with field theory, we conclude this section with the calculation of the gravitational free energy of our solution. This is inversely proportional to the effective $D=4$ Newton's constant, $F= \pi/(2 G_4)$ \cite{Emparan:1999pm}, which can be read off by inserting  (\ref{SU3U1lIIASolutionRescaled}) in the ten-dimensional action.  Denoting by $e^{2A}$ the warp factor in the metric of (\ref{SU3U1lIIASolutionRescaled}), and expressing the result in terms of $N$ and $k$, a straightforward calculation gives
\begin{eqnarray} \label{FreeEnergy1}
F= \frac{16 \pi^3}{(2\pi \ell_s )^8 }  \int_{S^6} e^{8A} \,  \textrm{vol}_6 = \frac{\pi}{5} \,   2^{1/3}  \, 3^{1/6} \, N^{5/3} \, k^{1/3} \; .
\end{eqnarray}

\noindent \textbf{Dual field theories}. We now ask whether there are large $N$ 3d conformal field theories dual to the AdS$_4$ massive IIA solutions obtained upon uplift of critical points of the ISO(7) supergravity. Because the internal manifold has the topology of $S^6$, it would be natural for these AdS$_4$ solutions to arise as the near horizon geometries of D2 branes in  smooth backgrounds of massive IIA. Such backgrounds, which must have curvature and RR and NS fields to be mutually supersymmetric with the D2 branes, have not been constructed. Nevertheless, we expect dual field theories with a single $\textrm{SU}(N)$ gauge group.

In flat space, the worldvolume theory of $N$ D2 branes in massless IIA is the maximally supersymmetric Yang-Mills theory in three dimensions with $\textrm{SU}(N)$ gauge group. It has 7 adjoint scalars and 8 fermions transforming under an $\textrm{SO}(7)$ R-symmetry. At low energies, this flows to the M2 brane conformal field theory with $\textrm{SO}(8)$ R-symmetry. On the Coulomb branch, the $N-1$ massless photons can be dualized in this three dimensional system to additional scalars, which complete the $\textrm{SO}(8)$ representation. Now, the presence of the Romans mass (\ref{QuantisationRomansMass}) induces a Chern-Simons term on the D2 brane, $\frac{k}{4\pi}\textrm{Tr} \left(A\wedge F + \frac{2}{3} A\wedge A \wedge A\right)$. By itself this would break all supersymmetry. However, we may take this together with additional couplings and preserve various numbers of supercharges, up to ${\cal N}=3$.

We will be more interested in the ${\cal N}=2$ Chern-Simons deformation. In ${\cal N}=2$ notation, the maximal 3d super Yang-Mills has an adjoint vector multiplet (containing a real scalar and a complex fermion) and 3 chiral multiplets (containing a complex scalar and fermion). There is a superpotential ${\cal W} = \textrm{Tr} X[Y, Z]$. The Chern-Simons deformation gives a mass to all fields in the vector multiplet. This leaves 6 real massless scalars. This theory has $\textrm{U}(1)_R \times \textrm{SU}(3)$ symmetry, like the massive IIA solution (\ref{SU3U1lIIASolutionRescaled}). The superpotential fixes the R-charge of the adjoint chiral multiplets to be $2/3$.

These field theories with a single gauge group and only adjoint matter are of the type explored by Schwarz in \cite{Schwarz:2004yj} as possible duals to AdS$_4$ string theory backgrounds. We will now give strong evidence that at least some of these simplest 3d theories are dual to the massive IIA uplifts of the AdS critical points of ISO(7) dyonic supergravity. We will match the gravitational free energy (\ref{FreeEnergy1}) of our solution (\ref{SU3U1lIIASolutionRescaled}) to the free energy of the $\cN=2$ Chern-Simons-matter theory that we have just described.

The gravitational free energy is dual to the 3d ${F = - \log Z}$, where $Z$ is the partition function of the SCFT on a Euclidean $S^3$. In general, $F$ is impossible to calculate in practice for strongly coupled field theories. However, ${\cal N}=2$ supersymmetry allows one to localize the infinite dimensional path integral to a finite dimensional integral over only supersymmetric configurations \cite{Kapustin:2009kz, Jafferis:2010un, Hama:2010av}.
The result is an integral, which is directly determined by the UV Lagrangian, over the eigenvalues of the adjoint scalar in the vector multiplet, i.e.~Coulomb branch parameters $\lambda_{i}$, \begin{eqnarray}Z = \int \prod_{i=1}^N \frac{d \lambda_i}{2\pi} \ \prod_{i<j=1}^N \left(2 \sinh^2(\frac{\lambda_i-\lambda_j}{2})\right) \times \nonumber \\ \prod_{i,j=1}^N \left(\exp(\ell(\frac{1}{3} + \frac{i}{2\pi}(\lambda_i-\lambda_j)))\right)^3 e^{\frac{i k}{4\pi} \sum \lambda_i^2},\end{eqnarray}
where $\sum \lambda_i=0$ since $su (N)$ is traceless. The $1/3$ that appears in the argument of $\ell(z) = -z \log (1-e^{2\pi i z}) + \frac{i}{2} \left( \pi z^2 + \frac{1}{\pi} \textrm{Li}_2 (e^{2\pi i z} ) \right) - \frac{i\pi}{12}$  results from the chiral multiplets having R-charge of $2/3$. This result is exact, even at finite $N$.

To compare to gravity, we want to take the large $N$ limit. Hence it is natural to describe the eigenvalues in terms of their density $\rho(\lambda)$. It is easy to check that the radius of curvature in string units of the string frame metric corresponding to (\ref{SU3U1lIIASolutionRescaled}) is of order $(\frac{N}{k})^{1/6}$, thus the supergravity solution is valid when $N \gg k$.  
In that limit, the range of $\lambda$ will scale as ${\lambda=N^\alpha (x+i y(x))}$, with $\alpha$ to be determined \cite{Herzog:2010hf, Jafferis:2011zi}. One finds an effective action for the eigenvalue density at large $N$
\begin{eqnarray}
S &=& \frac{N^{1+2\alpha}}{4 \pi} k \int d x \rho(x) \left(2 x y(x) - i (x^2 - y^2) \right)  \nonumber \\
&+& \frac{32}{27} \, \pi^2 \, N^{2-\alpha} \int d x \frac{\rho^2(x) }{1+y'(x)} \ .
\end{eqnarray}
The existence of a saddle point requires the two terms to balance, so $\alpha = 1/3$ as in \cite{Jafferis:2011zi}. The saddle point equation is algebraic, and one can easily find the solution. The result for the free energy is then
\begin{equation} F = \frac{3^{13/6} \pi}{40} \left(\frac{32}{27} \right)^{2/3} k^{1/3} N^{5/3}, \end{equation}
in exact agreement with the gravitational result (\ref{FreeEnergy1})! 

\vspace{2pt}

\noindent \textbf{Final comments}. This type of SCFTs with a simple gauge group and only adjoint matter has also been investigated in \cite{Minwalla:2011ma}. In fact, if we added a mass deformation $\mathrm{Tr} Z^2$, our $\cN=2$ theory would flow to the ${\cal N}=3$ point discussed there. We conjecture that this field theory is dual to the ${\cal N}=3$ AdS$_4 \times S^6$ massive IIA solution that arises from uplift via (\ref{KKEmbedding})--(\ref{KKmetric2}) of the $\cN=3$ point \cite{Gallerati:2014xra} of the dyonic ISO(7) theory. Interestingly, \cite{Minwalla:2011ma} demonstrated that there are light higher spin operators and an exponential growth in the spectrum of such simple theories with a larger number of adjoints. Thus the examples we have found appear to be the only possible such SCFTs with weakly curved supergravity duals.

Consistent truncations of $D=11$ and IIB supergravities on $S^n$ down to maximal supergravities have been extensively studied, with the usual $n=7,4,5$ cases singled out  \cite{Nastase:2000tu,Cvetic:2000dm,Lee:2014mla,Hohm:2014qga} as special. Now we can add a new consistent IIA, $n=6$ case. An interesting aspect in which the present $n=6$ case differs from  $n=7,4,5$ is that, although the resulting $D=4$ ISO(7) theory does not have an $\cN=8$ vacuum  that can possibly uplift to a(n in fact inexistent \cite{FigueroaO'Farrill:2002ft}) maximally supersymmetric ${\textrm{AdS}_4 \times S^6}$ type IIA background, a maximally supersymmetric  truncation does still exist at the level of the supergravities. Similarly, the question about the existence of a massive IIA truncation on $S^4$ to maximal $D=6$ supergravity could be addressed using the same approach. 

The dyonic deformation of the ISO(7) gauging is directly inherited from a deformation that already exists in the higher dimension. The embedding of $D=4$ dyonic gaugings in M-theory would require a different strategy, due to the absence of similar deformations of conventional $D=11$ supergravity.

\vspace{5pt}

\noindent \textbf{Acknowledgements}. We thank F.~Ciceri, B.~de Wit, C.~Hull, G.~Inverso, A.~Strominger, A.~Tomasiello and X.~Yin for helpful discussions. This work was supported in part by the Fundamental Laws Initiative at Harvard. A.G. is supported by the ERC Advanced Grant no. 246974, ``Supersymmetry: a window to non-perturbative physics''. The work of D.L.J. is supported in part by NSF CAREER grant PHY-1352084. OV is supported by the Marie Curie fellowship PIOF-GA-2012-328798, managed from the CPHT of \'Ecole Polytechnique.

\end{document}